# Analyzing Player Involvement in the Indian Pro Kabaddi League: A Network Analysis Approach

Arjab Sengupta[1[0009-0004-0641-6135]], Subhadip Layek[2[0009-0005-7004-7618]] and Krishanu Deyasi[3[0000-0002-0455-4495]]

[1] Department of Electronics and Communication Engineering
[2] Department of Electrical and Electronics Engineering
[3] Department of Basic Science and Humanities
Institute of Engineering & Management, Management House, D-1, Sector-V, Salt Lake Electronics Complex, Kolkata 700091, West Bengal, India
University of Engineering & Management, New Town, University Area, Plot No. III, B/5, New Town Rd, Action Area III, Newtown, Kolkata 700160, West Bengal, India

krishanu.deyasi@iem.edu.in

**Abstract.** This paper aims to apply network analysis to all players who have participated in the Indian Pro Kabaddi League since its inception. The Kabaddi network has been constructed based on the number of teams and players they have played with. The players have been ranked with the help of the degree and PageRank algorithm. Small-world phenomenon is observed in the Kabaddi network. The significance of the player's performance has been compared with the player's rank received by the network analysis.

## 1 Introduction

Social networks have become a fundamental paradigm for understanding the complexities of human interactions and relationships in various contexts [1, 2, 3]. With the advent of digital technologies and the proliferation of online platforms, studying social networks has gained significant attention across disciplines such as sociology, communication, and computer science [4, 5].

At its core, a social network comprises individuals or entities (nodes) interconnected by relationships or interactions (edges), forming a complex web of connections. These connections can manifest in diverse forms, including friendship network [6], coauthorship network [7], movie actor network [3], and communication network [8], sports network [9].

Network science tools have been applied in different sports to analyze the structure of the network. Peña, J.L. *et. al.* [10] used network theory to identify play patterns, potential weaknesses, and hotspots of the play using the passing data of the 2010 FIFA World

Cup. The best tennis player was identified by Radicchi, F. [11]. Fewell J.H. *et. al.* [12] used the basketball team as a network to quantify the winning strategy. By analyzing the cricket network, the performance of a player was determined by Mukherjee, S. [9]. Network theory applied to hockey by Stuart, H.C. [13]. In this paper, we delve into one new social network based on the Kabaddi sport.

Kabaddi is one of the most popular sports in India. As per the latest reports, an astonishing number of people watched Pro Kabaddi League season 10 [14]. Kabaddi is the national sport of Bangladesh. Kabaddi is not only popular in Asian countries but also in other countries around the world [15].

In this paper, we have constructed the Kabaddi network by grouping players based on their involvement in the team. We have applied network analysis techniques to investigate patterns of connectivity, clustering tendencies, and information flow dynamics. We used these criteria to rank the players and then compared the top-ranked players with their performance.

## 2 Network construction and structural analysis

We have gathered the data from the Kabaddi Adda website [16]. The Kabaddi network has been constructed using players who have participated in the Pro Kabaddi League (PKL). As of now, PKL has completed 10 seasons, spanning a decade of competition. The PKL was started in 2014 with 8 teams and currently, there are 12 teams playing in PKL [see Table 1]. The 12 teams are Patna Pirates, Puneri Paltan, U Mumba, Tamil Thalaivas, UP Yoddhas, Haryana Steelers, Dabang Delhi KC, Jaipur Pink Panthers, Bengal Warriors, Telugu Titans, Bengaluru Bulls and Gujarat Giants. Amongst these, Patna Pirates are the most successful franchise, having won three titles, more than any other franchise [16].

| | **1** | **2** | **3** | **4** | **5** | **6** | **7** | **8** | **9** | **10** |
|---|---|---|---|---|---|---|---|---|---|---|
| **Patna Pirates** | Y | Y | Y | Y | Y | Y | Y | Y | Y | Y |
| **Puneri Paltan** | Y | Y | Y | Y | Y | Y | Y | Y | Y | Y |
| **U Mumba** | Y | Y | Y | Y | Y | Y | Y | Y | Y | Y |
| **Tamil Thalaivas** | | | | | Y | Y | Y | Y | Y | Y |
| **UP Yoddhas** | | | | | Y | Y | Y | Y | Y | Y |
| **Haryana Steelers** | | | | | Y | Y | Y | Y | Y | Y |
| **Dabang Delhi KC** | Y | Y | Y | Y | Y | Y | Y | Y | Y | Y |
| **Jaipur Pink Panthers** | Y | Y | Y | Y | Y | Y | Y | Y | Y | Y |
| **Bengal Warriors** | Y | Y | Y | Y | Y | Y | Y | Y | Y | Y |
| **Telugu Titans** | Y | Y | Y | Y | Y | Y | Y | Y | Y | Y |
| **Bengaluru Bulls** | Y | Y | Y | Y | Y | Y | Y | Y | Y | Y |
| **Gujarat Giants** | | | | | Y | Y | Y | Y | Y | Y |

**Table 1:** Here are all the 12 teams in the Pro Kabaddi League and their participation in the last 10 seasons.

Two Kabaddi players are assumed to be connected by an edge if they are part of the same squad in the same season. For example, if players A, B, and C played together on a team in one season, and in another season players A, C, D and E were teammates, then graphically they are represented by Figure 1.

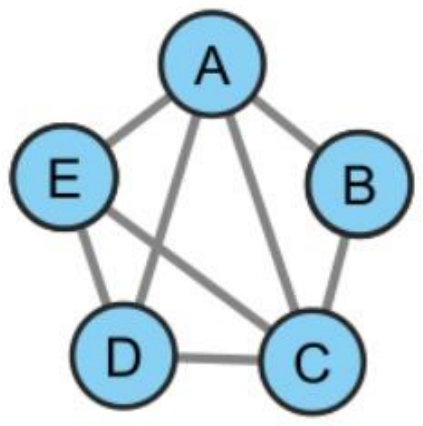

**Figure 1**: Sample network with 5 players. In one season, players A, B, and C were on the same team, while in another season, players A, C, D and E were on the same team.

The constructed network has 863 nodes and 17195 edges. This means that, on average, each player has played around 19 other players in their PKL career. We found that all nodes of the network is in a single giant component [see Figure 2]. This signifies that all the players are connected. In the next section, *degree*, *clustering coefficient* and *average shortest distance* have been calculated. The clustering coefficient and average shortest distance of the Kabaddi network have been compared with the two model networks, *viz.*, Erdös–Rényi (ER) network [17] and the configuration network [18]. PageRank analysis has also been conducted to rank the players.

**2.1 Degree:** Degree of a node *i* in a network is defined as [2]:

$$d_i = \sum_{j=1}^{N} A_{ij}, \qquad (1)$$

where A is called the adjacency matrix defined as $A_{ij} = 1$ if the players i and j are in the same squad and $A_{ij} = 0$ if the players i and j are in the different squads, N is the total number of players in the network.

From our analysis, PO Surjeet Singh stands out as the player with the highest degree. He has played with 159 players. The top 40 players with the highest degrees are listed in Table 2.

| **Degree Rank** | **Player name** | **Degree ($d_i$)** |
|---|---|---|
| 1 | PO SURJEET SINGH | 159 |
| 2 | GIRISH MARUTI ERNAK | 154 |
| 3 | K PRAPANJAN | 151 |
| 4 | RAKESH NARWAL | 144 |
| 5 | RAVI KUMAR | 144 |
| 6 | RAHUL CHAUDHARI | 143 |
| 7 | PRASHANTH KUMAR RAI | 143 |
| 8 | ASISH KUMAR SANGWAN | 138 |
| 9 | CHANDRAN RANJIT | 138 |
| 10 | RAVINDER PAHAL | 137 |
| 11 | SELVAMANI K | 136 |
| 12 | DEEPAK NIWAS HOODA | 134 |
| 13 | FAZEL ATRACHALI | 131 |
| 14 | VIJIN THANGADURAI | 131 |
| 15 | SANDEEP NARWAL | 130 |
| 16 | ANIL KUMAR | 126 |
| 17 | PAWAM KUMAR SEHRAWAT | 125 |
| 18 | AMIT HOODA | 125 |
| 19 | PARDEEP NARWAL | 125 |
| 20 | SHRIKANT JADHAV | 123 |
| 21 | VISHAL PRABHAKAR MANE | 122 |
| 22 | SUKESH HEGDE | 121 |
| 23 | AJAY THAKUR | 118 |
| 24 | RAN SINGH | 118 |
| 25 | NITIN TOMAR | 117 |
| 26 | DHARMARAJ CHERALATHAN | 116 |
| 27 | DEEPAK NARWAL | 116 |
| 28 | SURENDER NADA | 116 |
| 29 | VIKAS KANDOLA | 115 |
| 30 | MOHIT CHHILLAR | 115 |
| 31 | HADI OSHTORAK | 115 |
| 32 | MANJEET CHHILLAR | 114 |
| 33 | MAHENDRA GANESH RAJPUT | 113 |
| 34 | SANDEEP DHULL | 111 |
| 35 | VISHAL BHARADWAJ | 111 |
| 36 | PAWAN KUMAR KADIYAN | 110 |
| 37 | PARVESH BHAINSWAL | 110 |
| 38 | RAJESH NARWAL | 109 |
| 39 | C ARUN | 109 |
| 40 | MONU GOYAT | 108 |

**Table 2**: Top 40 highly connected players in the Kabaddi network.

**2.2 Clustering coefficient:** The clustering coefficient ($C_i$) of a node i is defined as the ratio of the number of edges shared by its neighboring nodes to the maximum number of possible edges among them [2]. Hence, the average clustering coefficient of a network is defined as:

$$C = \frac{1}{N}\sum_{i=1}^{N} C_i \qquad (2)$$

We have got the clustering coefficient of the Kabaddi network to be 0.7280 signifying that the Kabaddi network is highly clustered. The ER network on the other hand has a clustering coefficient of about 0.0462, *i.e.*, in an order that is lower in magnitude than the Kabaddi network. The configuration network has been generated from the original network by keeping the same number of nodes and the same degree sequence as that of the original network. The clustering coefficient of the configuration network is 0.0927, much less than the Kabaddi network but higher than the ER network.

**2.3 Average shortest distance:** The average shortest distance *l* between any two vertices is defined as [2]:

$$l = \frac{2}{N(N-1)}\sum_{i,j \in V} d\,(i,j) \qquad (3)$$

where $d(i, j)$ denotes the minimum number of edges required to reach from vertex *i* to *j* and V is the set of vertices.

The average shortest path distance of the Kabaddi network is 2.349. The high clustering coefficient (C) and short average distance (*l*) indicate that the Kabaddi network is a small world network [3]. Compared to the Kabaddi network, the average shortest distance of the ER and configuration networks are 2.105 and 2.187, respectively.

| Network | Number of nodes (N) | Number of edges (E) | Average clustering coef-ficient (C) | Average shortest distance (l) |
|---|---|---|---|---|
| Kabaddi net-work (PKL) | 863 | 17195 | 0.7280 | 2.349 |
| Erdös–Rényi model | 863 | 17195 | 0.0462 | 2.105 |
| Configuration model | 863 | 17195 | 0.0927 | 2.187 |

**Table 2:** Network statistics of the Kabaddi network and the corresponding Erdös–Rényi model [17] and configuration model [18].

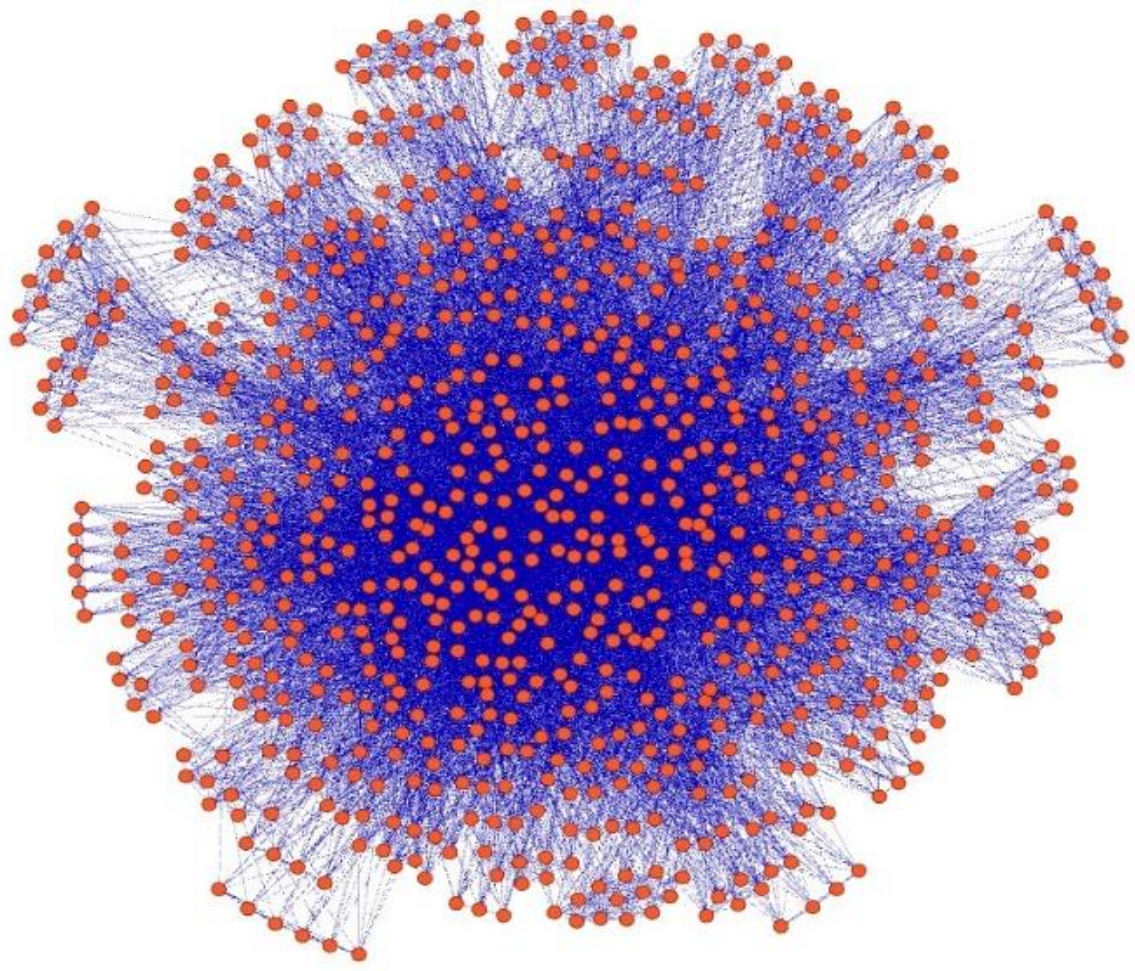

**Figure 2**: Network structure of players in the Pro Kabaddi League.

## 3 PageRank analysis

In this section, players have been sorted out in descending order based on the PageRank score of a particular node. PageRank was originally developed by Sergey Brin and Larry Page in 1998 [19]. PageRank is used to rank web pages to deliver search results relevant to user queries. The web pages are considered as nodes and the corresponding hyperlinks which lead to that particular page are considered as edges. The PageRank for an undirected network is defined as [4]:

$$Rank(u_i) = \frac{1-c}{N} + c\sum_{j=1}^{N} A_{ij}\frac{Rank(u_j)}{k_j} \qquad (4)$$

Here, c is the damping factor/jump factor. Typically, the value of $c$ is set to 0.85. Rank ($u_i$) and Rank ($u_j$) are the PageRank scores of the vertices i and j, respectively. $A_{ij}$ is the ijth element of the adjacency matrix A. $k_j$ is the degree of the vertex j.

In the PageRank analysis, Girish Maruti Ernak emerges as the top-ranked player in the network. Table 3 presents the top 40 players from all 10 seasons of the Pro Kabaddi League as determined by the PageRank analysis.

We have taken the average strike rates of the individual player from [16] and compared them with the PageRank scores of a particular player/node. The average strike rate of a player quantifies the performance of a player over the years. Based on the PageRank score, the top 40 players have been grouped together depending on their assigned scores in Table 3. PageRank score and average strike rate of the top 40 players is negatively correlated with the correlation coefficient value -0.5684. A higher PageRank score indicates a greater number of edges, and therefore, a higher degree [20]. This means that the particular player has played with a relatively larger number of teams compared to the rest of the players. This suggests that the particular player has been released by the franchise or team before the auction of a particular season, indicating that the player's performance within the team may not have been satisfactory for retention. Naturally, we observe an inverse relationship between the overall strike rates of the players and their PageRank scores.

| **Rank** | **Name** | **PageRank Score** | **Avg Strike Rate (%)** |
|---|---|---|---|
| 1 | GIRISH MARUTI ERNAK | 0.003996 | 45.12 |
| 2 | K PRAPANJAN | 0.003991 | 42.21 |
| 3 | PO SURJEET SINGH | 0.003959 | 51.96 |
| 4 | RAHUL CHAUDHARI | 0.003778 | 45.84 |
| 5 | RAKESH NARWAL | 0.003700 | 54.31 |
| 6 | RAVI KUMAR | 0.003608 | 52.98 |
| 7 | SELVAMANI K | 0.003586 | 46.80 |
| 8 | ASISH KUMAR SANGWAN | 0.003584 | 48.44 |
| 9 | CHANDRAN RANJIT | 0.003566 | 40.29 |
| 10 | PRASHANTH KUMAR RAI | 0.003537 | 49.90 |
| 11 | RAVINDER PAHAL | 0.003497 | 51.07 |
| 12 | PAWAN KUMAR SEHRAWAT | 0.003384 | 66.52 |
| 13 | FAZEL ATRACHALI | 0.003361 | 52.79 |
| 14 | SANDEEP NARWAL | 0.003331 | 39.65 |
| 15 | DEEPAK NIWAS HOODA | 0.003316 | 48.02 |
| 16 | VIJIN THANGADURAI | 0.003291 | 44.92 |
| 17 | ANIL KUMAR | 0.003241 | 57.36 |
| 18 | AMIT HOODA | 0.003204 | 53.81 |
| 19 | PARDEEP NARWAL | 0.003190 | 54.34 |
| 20 | SHRIKANT JADHAV | 0.003170 | 47.80 |
| 21 | VISHAL PRABHAKAR MANE | 0.003102 | 52.72 |
| 22 | SUKESH HEGDE | 0.003006 | 44.84 |
| 23 | NITIN TOMAR | 0.002992 | 50.13 |
| 24 | AJAY THAKUR | 0.002992 | 45.95 |
| 25 | VISHAL BHARADWAJ | 0.002980 | 57.39 |
| 26 | RAN SINGH | 0.002970 | 49.50 |
| 27 | SANDEEP DHULL | 0.002956 | 52.25 |
| 28 | DHARMARAJ CHARALATHAN | 0.002935 | 54.98 |
| 29 | MAHENDRA GANESH RAJPUT | 0.002919 | 47.43 |
| 30 | VIKAS KANDOLA | 0.002917 | 48.88 |
| 31 | DEEPAK NARWAL | 0.002915 | 45.63 |
| 32 | SURENDER NADA | 0.002896 | 45.22 |
| 33 | MANJEET CHHILLAR | 0.002893 | 53.63 |
| 34 | PARVESH BHAINSWAL | 0.002886 | 60.01 |
| 35 | MOHIT CHHILLAR | 0.002874 | 47.27 |
| 36 | SANTHAPANASELVAM | 0.002873 | 55.40 |
| 37 | SOMBIR | 0.002872 | 55.97 |
| 38 | HADI OSHTORAK | 0.002864 | 55.90 |
| 39 | C ARUN | 0.002819 | 53.31 |
| 40 | SHRIKANT TEWTHIA | 0.002797 | 51.37 |

**Table 3**: Top 40 high PageRank players in the Kabaddi network.

## 4 Discussion and conclusion

In this paper, we have constructed the Kabaddi network based on the participation of players in the Pro Kabaddi League. We have examined the Kabaddi network as a simple graph, with players as nodes, and an edge connecting two nodes if those players have been on the same team. Our analysis reveals that the graph has 863 nodes and 17195 edges. It has been observed that the Kabaddi network is highly clustered with the clustering coefficient of 0.7280 and the average shortest path length *l* which comes out to be 2.349. Consequently, the network possesses small-world properties. The average degree of the constructed network is 39.83. Finally, we determine the PageRank score of the network. The top 40 players, based on their PageRank scores, unsurprisingly display an inverse relationship with their average strike rates. In some cases, the strike rates of the players tended to drop amongst players who have played the same number of matches. The reason is the players' underperformance, leading to multiple releases from their teams. Consequently, they have played with more teammates, increasing their degrees. In some cases, higher strike rates accompanying higher PageRank scores can be attributed to players consistently performing well. Teams show interest in recruiting these players for future seasons, offering them more money, leading to players switching teams.

## References

1. Wasserman, S. and Faust, K., 1994. Social network analysis: Methods and applications.
2. Newman, M.E., 2003. The structure and function of complex networks. *SIAM review*, *45*(2), pp.167-256.
3. Watts, D.J. and Strogatz, S.H., 1998. Collective dynamics of ‘small-world’networks. *Nature*, *393*(6684), pp.440-442.
4. Newman, M., 2018. *Networks*. Oxford University Press.
5. Albert, R. and Barabási, A.L., 2002. Statistical mechanics of complex networks. *Reviews of modern physics*, *74*(1), p.47.
6. Leskovec, J. and Mcauley, J., 2012. Learning to discover social circles in ego networks. *Advances in neural information processing systems*, *25*.
7. Newman, M.E., 2006. Finding community structure in networks using the eigenvectors of matrices. *Physical review E*, *74*(3), p.036104.
8. Guimera, R., Danon, L., Diaz-Guilera, A., Giralt, F. and Arenas, A., 2003. Self-similar community structure in a network of human interactions. *Physical Review E*, *68*(6), p.065103.
9. Mukherjee, S., 2014. Quantifying individual performance in Cricket—A network analysis of Batsmen and Bowlers. *Physica A: Statistical Mechanics and its Applications*, *393*, pp.624-637.
10. Peña, J.L. and Touchette, H., 2012. A network theory analysis of football strategies. *arXiv preprint arXiv:1206.6904*.
11. Radicchi, F., 2011. Who is the best player ever? A complex network analysis of the history of professional tennis. *PloS one*, *6*(2), p.e17249.
12. Fewell, J.H., Armbruster, D., Ingraham, J., Petersen, A. and Waters, J.S., 2012. Basketball teams as strategic networks. *PloS one*, *7*(11), p.e47445.

13. Stuart, H.C., 2017. Structural disruption, relational experimentation, and performance in professional hockey teams: A network perspective on member change. *Organization Science*, *28*(2), pp.283-300.
14. https://www.hindustantimes.com/sports/others/pro-kabaddi-league-season-10-records-226-million-viewers-in-first-90-matches-101706806411039.html
15. https://www.kabaddiikf.com/members-country.html
16. https://www.kabaddiadda.com/
17. Erdős, P. and Rényi, A., 1960. On the evolution of random graphs. *Publ. math. inst. hung. acad. sci*, *5*(1), pp.17-60.
18. Newman, M.E., Strogatz, S.H. and Watts, D.J., 2001. Random graphs with arbitrary degree distributions and their applications. *Physical Review E*, *64*(2), p.026118.
19. Brin, S. and Page, L., 1998. The anatomy of a large-scale hypertextual web search engine. *Computer networks and ISDN systems*, *30*(1-7), pp.107-117.
20. Deyasi, K., 2016. On the initial value of PageRank. *arXiv preprint arXiv:1609.00004*.